\documentclass{ifacconf}
\usepackage{float}
\usepackage{amsmath} 
\usepackage{amssymb}
\usepackage{algorithm}
\usepackage{algpseudocode}
\usepackage{booktabs} 
\usepackage{makecell}
\usepackage{afterpage} 
\usepackage{graphicx}      
\usepackage{natbib}        
\begin{document}
\begin{frontmatter}

\title{Stability-Guaranteed Dual Kalman Filtering for Electrochemical Battery State Estimation\thanksref{footnoteinfo}} 
\thanks[footnoteinfo]{This work has been submitted to IFAC for possible publication.}

\author[First,Second,Third]{Feng Guo} 
\author[Second]{Guangdi Hu} 
\author[Fifth]{Keyi Liao}
\author[Second,Third]{Luis D. Couto} 
\author[Second,Third]{Khiem Trad} 
\author[Sixth]{Ru Hong} 
\author[First,Third,Seventh]{Hamid Hamed} 
\author[First,Third,Seventh]{Mohammadhosein Safari}

\address[First]{Institute for Materials Research (IMO-imomec),Hasselt University (UHasselt), 3500, Hasselt, Belgium (e-mail: \{feng.guo, hamid.hamed, momo.safari\}@uhasselt.be).}
\address[Second]{VITO, 2400, Mol, Belgium (e-mail: \{luis.coutomendonca, khiem.trad\}@vito.be)}
\address[Third]{EnergyVille, 3600, Genk, Belgium }
\address[Fourth]{School of Vehicles and Intelligent Transportation, Fuyao University of Science and Technology, Fuzhou, China (e-mail: guangdihu@fyust.edu.cn)}
\address[Fifth]{School of Mechanical Engineering, Southwest Jiaotong University, Chengdu, China, (e-mail: 2018210514@my.swjtu.edn.cn)}
\address[Sixth]{College of Tianyin Lake Science and Technology Innovation, Nanjing Institute of Technology, 211167, Nanjing, China (e-mail: ruhong\_ai@163.com )}
\address[Seventh]{IMEC Division IMOMEC, 3590, Diepenbeek, Belgium}

\begin{abstract}
Accurate and stable state estimation is critical for battery management. Although dual Kalman filtering can jointly estimate states and parameters, the strong coupling between filters may cause divergence under large initialization errors or model mismatch. This paper proposes a Stability Guaranteed Dual Kalman Filtering (SG-DKF) method. A Lyapunov-based analysis yields a sufficient stability condition, leading to an adaptive dead-zone rule that suspends parameter updates when the innovation exceeds a stability bound. Applied to an electrochemical battery model, SG-DKF achieves accuracy comparable to a dual EKF and reduces state of charge RMSE by over 45\% under large initial state errors.
\end{abstract}

\begin{keyword}
 Nonlinear Adaptive Control; Kalman Filtering; Lyapunov Methods; Stability of Nonlinear Systems; Energy Management Systems; Chemical Process Control
\end{keyword}

\end{frontmatter}

\section{Introduction}
Accurate estimation of internal states in lithium-ion batteries is essential for guaranteeing safety, efficiency, and optimal operation in electric vehicles and stationary energy storage systems~\citep{safari2025perspective,haghverdi2025review}. Electrochemical battery models provide a physically interpretable representation of diffusion and reaction dynamics, and their states such as concentration variables and state of charge cannot be measured directly~\citep{guo2024systematic}. Kalman filtering-based estimation strategies have therefore become a central component of battery management algorithms~\citep{plett2004extended}. Among these strategies, dual Kalman filtering is frequently adopted because it provides a practical way to estimate both model states and time varying parameters~\citep{wan2001dual}. As battery systems continue to operate under wider dynamic conditions and stronger uncertainty, developing state estimation methods that remain reliable for electrochemical models is increasingly important.

Battery modeling approaches commonly fall into two major categories: equivalent circuit models (ECM) and electrochemical models. ECM approximate the terminal voltage response using combinations of voltage sources, resistors, and capacitors~\citep{plett2004extended2}. Their main advantage lies in their simple structure and computational efficiency, which makes them attractive for embedded battery management systems. However, because these models do not capture the underlying electrochemical processes, they offer limited insight into internal variables and are difficult to use for advanced estimation and control tasks under stringent safety and performance requirements. Electrochemical models, on the other hand, describe battery behavior based on fundamental transport and reaction mechanisms. The most representative formulation is the pseudo-two-dimensional (P2D) model~\citep{fuller1994simulation}, together with reduced-order models derived from it, such as the single particle model (SPM)~\citep{haran1998determination,guo2025comparative,guo2025control} and  SPM with electrolyte~\citep{marquis2019asymptotic}. These models are capable of reflecting internal diffusion dynamics and providing access to quantities that are otherwise unmeasurable, which makes them highly suitable for state estimation and predictive control. Nevertheless, full electrochemical models involve a large number of parameters and exhibit significant nonlinearities, which pose challenges for real-time implementation. As a result, simplified variants, particularly the SPM and its extensions, are widely adopted in practical battery management applications~\citep{guo2024systematic}.

Kalman filtering has become a cornerstone technique for battery state estimation due to its ability to fuse model predictions with noisy measurements in a recursive and computationally efficient manner~\citep{plett2004extended3}. Dual Kalman filtering extends this framework by employing one filter to estimate the system states and another filter to estimate time-varying model parameters. This separation often leads to improved estimation accuracy compared with a single Kalman filter, because the online parameter estimation partially compensates for modeling errors and  parameter variations. Recent studies have applied dual Kalman filtering to electrochemical battery models and have demonstrated promising performance in estimating both internal states and parameters in real time~\citep{gao2021co,hosseininasab2023state}. However, because the two filters are tightly coupled, an increase in estimation error in either the state filter or the parameter filter can propagate to the other, which may eventually cause divergence or instability of the overall estimation process~\citep{slotine1991applied}.

Extensive research has been conducted on the stability of single Kalman filter. For example, \cite{zhang2026stability} analyzed the stability of the Kalman filter under practical conditions, while \cite{li2012stochastic} investigated the stochastic stability of the unscented Kalman filter with intermittent observations. In contrast, the stability of dual Kalman filtering has received far less attention, and systematic theoretical results are still lacking. In our previous work, we introduced a dead-zone-based dual Kalman filtering strategy for ECM to alleviate potential instability~\citep{guo2018parameter}. The idea was to suspend the update of the parameter filter when the model mismatch exceeded a predefined threshold, thereby preventing error propagation and improving numerical robustness. However, that approach did not provide a rigorous stability proof.

Building on this idea, the present study incorporates Lyapunov-based analysis to establish a stability-guaranteed dual Kalman filtering (SG-DKF) framework, and the method is developed using a more accurate electrochemical battery model. This allows us to formalize the stability mechanism of dual Kalman filtering and provide theoretical conditions under which the coupled estimation process remains bounded.

\section{Stability-Guaranteed Dual Kalman Filtering}
\label{sec:SGDKF}

This section summarizes the dual extended Kalman filtering (Dual EKF)
framework and the stability-guaranteed dead-zone mechanism. The dual
structure consists of (i) a \emph{state EKF} estimating the fast
electrochemical states, and (ii) a \emph{parameter EKF} estimating
slowly varying model parameters \citep{wan2001dual}.
\subsection{Dual EKF Structure}
Consider the nonlinear discrete-time system
\begin{equation}
x_k = f(x_{k-1},\theta_{k-1},u_{k-1}) + w_{k-1}, 
y_k = h(x_k,\theta_{k-1},u_k) + v_k,
\label{eq:nonlinear_sys}
\end{equation}
where $x_k \in \mathbb{R}^n$ is the state, $\theta_k \in \mathbb{R}^m$
is the parameter vector, $u_k$ is the input, $y_k$ is the measured
output, and $w_k$, $v_k$ denote process and measurement noise.
Linearizing \eqref{eq:nonlinear_sys} along the EKF trajectory gives
the LTI approximation
\begin{equation}
x_{k+1} = A x_k + B u_k + w_k, \qquad
y_k = C x_k + D u_k + v_k,
\label{eq:lti_model}
\end{equation}
where $A$ and $C$ are the Jacobians of $f$ and $h$ with respect to
$x$, and $B$, $D$ are the input matrices. 

For State EKF. Given $\hat{\theta}_{k-1|k-1}$, the state EKF performs the standard
prediction–update steps \citep{plett2004extended3}:
\begin{align}
\hat{x}_{k|k-1} &= f(\hat{x}_{k-1|k-1},\hat{\theta}_{k-1|k-1},u_{k-1}),\\
P_{k|k-1} &= A_{k-1} P_{k-1|k-1} A_{k-1}^\top + Q, \\
K_k &= P_{k|k-1} C_k^\top \bigl(C_k P_{k|k-1} C_k^\top + R\bigr)^{-1},\\
E_k &= y_k - h(\hat{x}_{k|k-1},\hat{\theta}_{k-1|k-1},u_k),\\
\hat{x}_{k|k} &= \hat{x}_{k|k-1} + K_k E_k,\\
P_{k|k} &= (I - K_k C_k) P_{k|k-1},
\end{align}
where $A_{k-1} = \partial f / \partial x|_{\hat{x}_{k-1|k-1}}$ and
$C_k = \partial h / \partial x|_{\hat{x}_{k|k-1}}$.

For parameter EKF.
The parameters are modelled as a random walk,
\begin{equation}
\theta_k = \theta_{k-1} + \omega_{k-1},
\end{equation}
with process noise $\omega_k$. The parameter EKF uses the latest state
estimate $\hat{x}_{k|k}$:
\begin{align}
\hat{\theta}_{k|k-1} &= \hat{\theta}_{k-1|k-1}, \\
P_{\theta,k|k-1} &= P_{\theta,k-1|k-1} + Q_\theta, \\
K_{\theta,k} &=
  P_{\theta,k|k-1} C_{\theta,k}^\top
  \bigl(C_{\theta,k} P_{\theta,k|k-1} C_{\theta,k}^\top + R\bigr)^{-1},\\
r_k &= y_k - h_p(\hat{\theta}_{k|k-1},\hat{x}_{k|k},u_k),\\
\hat{\theta}_{k|k} &= \hat{\theta}_{k|k-1} + K_{\theta,k} r_k,\\
P_{\theta,k|k} &= (I - K_{\theta,k} C_{\theta,k}) P_{\theta,k|k-1},
\end{align}
where $h_p(\cdot)$ is the output mapping with respect to $\theta$, and
$C_{\theta,k} = \partial h_p / \partial \theta|_{\hat{\theta}_{k|k-1}}$.

\subsection{Stability-Guaranteed Dead-Zone Criterion}

Because the two EKFs are coupled, simultaneous adaptation can cause
divergence under large innovation or model mismatch. To ensure
stability, we analyze the local error dynamics of the state EKF and
use the result to define a dead zone for the
parameter EKF.

Let $\hat{x}_{k|k}$ be the state estimate and define
\begin{equation}
\tilde{x}_k = x_k - \hat{x}_{k|k}.
\end{equation}
To eliminate the explicit input dependence in the state equation,
introduce the shifted variables
\begin{equation}
\hat{z}_k = \hat{x}_{k|k} + A^{-1} B u_{k-1}, 
\tilde{z}_k = x_k + A^{-1} B u_{k-1} - \hat{z}_k.
\end{equation}
Since $\tilde{z}_k$ and $\tilde{x}_k$ differ only by a known quantity,
boundedness of one implies boundedness of the other.

Using the prediction–update form of the state EKF and collecting
terms, the error dynamics can be written compactly as
\begin{equation}
\tilde{z}_{k+1} = A \tilde{z}_k + \tilde{w}_k,
\label{eq:error_dynamics}
\end{equation}
with an effective disturbance
\begin{equation}
\tilde{w}_k = w_k + K_{k+1} E_{k+1},
\label{eq:w_tilde_def}
\end{equation}
where $K_{k+1}$ is the Kalman gain and
\begin{equation}
E_{k+1}
= y_{k+1}
  - \bigl(C \hat{x}_{k+1|k} + D u_{k+1}\bigr)
\label{eq:innovation_def}
\end{equation}
is the innovation of the state EKF in the linearized model. Because
the true output $y_{k+1}$ and the predicted output share the same
input $u_{k+1}$, the direct input term $D u_{k+1}$ cancels out in the
error dynamics and does not affect \eqref{eq:error_dynamics}.

For Lyapunov inequality.
Assume that $A$ is Schur. Then, for any $Q = Q^\top > 0$, the
discrete-time Lyapunov equation
\begin{equation}
A^\top P A - P = -Q
\end{equation}
admits a unique solution $P = P^\top > 0$. Consider the Lyapunov
function
\begin{equation}
V_k = \tilde{z}_k^\top P \tilde{z}_k.
\end{equation}
Using \eqref{eq:error_dynamics}, the one-step increment is
\begin{align}
\Delta V_{k+1}
&:= V_{k+1} - V_k \nonumber\\
&= (A\tilde{z}_k + \tilde{w}_k)^\top
    P (A\tilde{z}_k + \tilde{w}_k)
   - \tilde{z}_k^\top P \tilde{z}_k \nonumber\\
&= \tilde{z}_k^\top (A^\top P A - P) \tilde{z}_k
   + 2 \tilde{z}_k^\top A^\top P \tilde{w}_k
   + \tilde{w}_k^\top P \tilde{w}_k \nonumber\\
&= - \tilde{z}_k^\top Q \tilde{z}_k
   + 2 \tilde{z}_k^\top A^\top P \tilde{w}_k
   + \tilde{w}_k^\top P \tilde{w}_k.
\end{align}
To bound the cross term we use Young’s inequality
\citep[Lemma~1]{khalil2002nonlinear}. Let
$a = P^{1/2} A \tilde{z}_k$ and $b = P^{1/2} \tilde{w}_k$. Then, for
any $\varepsilon>0$,
\begin{align}
2 \tilde{z}_k^\top A^\top P \tilde{w}_k
 &= 2 a^\top b \nonumber\\
 &\le \varepsilon \|a\|^2
    + \frac{1}{\varepsilon}\|b\|^2 \nonumber\\
 &= \varepsilon\, \tilde{z}_k^\top A^\top P A\, \tilde{z}_k
    + \frac{1}{\varepsilon}\, \tilde{w}_k^\top P\, \tilde{w}_k.
\label{eq:young_bound}
\end{align}
Using the bounds
\begin{align}
\lambda_{\min}(Q)\,\|\tilde{z}_k\|^2
  &\le \tilde{z}_k^\top Q \tilde{z}_k
   \le \lambda_{\max}(Q)\,\|\tilde{z}_k\|^2,
\label{eq:Q_bound}\\[3pt]
\tilde{z}_k^\top A^\top P A\, \tilde{z}_k
  &\le \|A^\top P A\|\,\|\tilde{z}_k\|^2.
\label{eq:APA_bound}
\end{align}
and $\tilde{w}_k^\top P \tilde{w}_k \le \|P\|\,\|\tilde{w}_k\|^2$, we
obtain
\begin{align}
\Delta V_{k+1}
&\le
\Bigl(
 -\lambda_{\min}(Q)
 + \varepsilon \|A^\top P A\|
\Bigr)\|\tilde{z}_k\|^2 \nonumber\\
&\quad
+ \left(1 + \frac{1}{\varepsilon}\right)\|P\|\,\|\tilde{w}_k\|^2.
\end{align}
Choose $\varepsilon>0$ small enough such that
\begin{equation}
\alpha := \lambda_{\min}(Q) - \varepsilon \|A^\top P A\| > 0,
\end{equation}
and define
\begin{equation}
\beta := \left(1 + \frac{1}{\varepsilon}\right)\|P\|.
\end{equation}
Then
\begin{equation}
\Delta V_{k+1}
\le -\alpha \|\tilde{z}_k\|^2
     + \beta \|\tilde{w}_k\|^2.
\label{eq:deltaV_bound_alpha_beta}
\end{equation}
Thus $V_k$ is strictly decreasing whenever
\begin{equation}
\|\tilde{w}_k\|^2
<
\frac{\alpha}{\beta}\,\|\tilde{z}_k\|^2
\;\;\Rightarrow\;\;
\|\tilde{w}_k\|^2
<
\frac{\lambda_{\min}(Q)}{\|I+P\|}\,\|\tilde{z}_k\|^2,
\label{eq:stability_condition}
\end{equation}
which provides a sufficient local stability condition for the state
EKF in the presence of the disturbance \eqref{eq:w_tilde_def}.

Since the exact norms of $\tilde{z}_k$ and $w_k$ are not available,
they are conservatively bounded via the state covariance $P_{k|k}$ and
the process noise covariance $Q$. This leads to an implementable
innovation threshold
\begin{equation}
\delta_k =
\frac{
\sqrt{\lambda_{\min}(Q)/\|I+P\|}
\;\|\tilde{z}_k\| + \|w_k\|
}{\|K_{k+1}\|},
\label{eq:dead_zone_final}
\end{equation}
which is consistent with \eqref{eq:stability_condition} after
bounding the constants. The dead-zone switching signal is defined as
\begin{equation}
\sigma_k =
\begin{cases}
1, & \|E_k\| < \delta_k,\\
0, & \|E_k\| \ge \delta_k,
\end{cases}
\label{eq:switch}
\end{equation}
where $\sigma_k = 1$ allows the parameter EKF to update, and
$\sigma_k = 0$ freezes the parameter estimate to avoid destabilizing
the state filter. This innovation-based dead zone is the core
mechanism that guarantees the local stability of the proposed SG-DKF.

\subsection{Overall SG-DKF Procedure}

At each sampling step, the SG-DKF executes: (i) a full state EKF step,
(ii) computation of the innovation and admissible bound $\delta_k$, and
(iii) parameter update only when $\sigma_k = 1$. The overall procedure
is summarized in Algorithm~\ref{alg:SGDKF}.

\begin{algorithm}[h]
\caption{Stability-Guaranteed Dual Kalman Filtering}
\label{alg:SGDKF} 
\begin{algorithmic}[1]
\State Initialize $\hat{x}_{0|0}$, $P_{0|0}$, $\hat{\theta}_{0|0}$, $P_{\theta,0|0}$
\For{$k = 1,2,\ldots$}
    \State State EKF prediction and update
    \State Compute innovation $E_k$ and threshold $\delta_k$
    \State Compute switching signal $\sigma_k$ via \eqref{eq:switch}
    \If{$\sigma_k = 1$}
        \State Parameter EKF prediction and update
    \Else
        \State Freeze $\hat{\theta}_{k|k}$ and $P_{\theta,k|k}$
    \EndIf
\EndFor
\end{algorithmic}
\end{algorithm}

\begin{figure*}[h]
    \centering
    \includegraphics[width=0.825\textwidth]{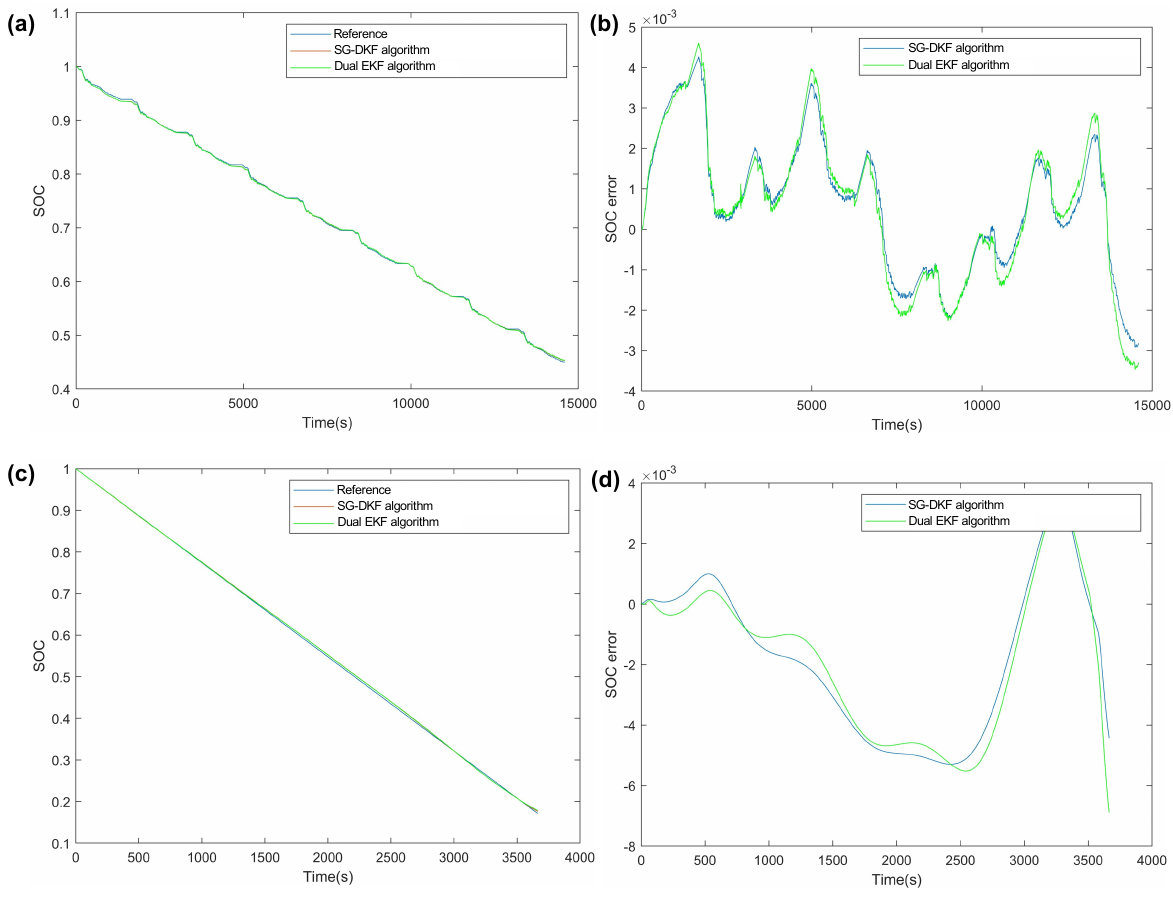}
\caption{SOC estimation performance without initial SOC error for the dual EKF and SG-DKF under UDDS and 1C discharge:
(a) SOC under UDDS;
(b) SOC error under UDDS;
(c) SOC under 1C;
(d) SOC error under 1C.}
    \label{fig:1}
\end{figure*}

\begin{figure*}[h]
    \centering
    \includegraphics[width=0.825\textwidth]{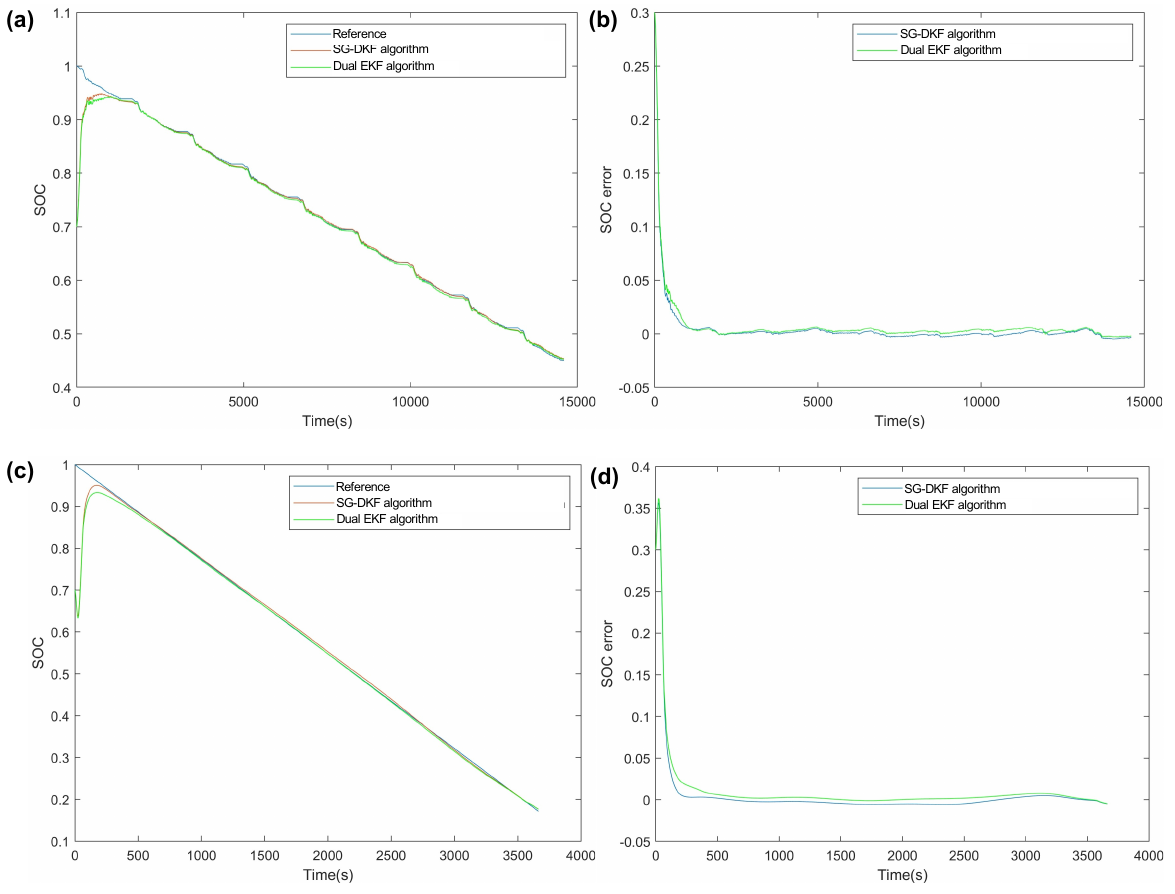}
\caption{SOC estimation performance with 30\% initial SOC error for the dual EKF and the SG-DKF under UDDS and 1C discharge: 
(a) SOC under UDDS; 
(b) SOC error under UDDS; 
(c) SOC under 1C; 
(d) SOC error under 1C.}
    \label{fig:2}
\end{figure*}

\section{Battery Model}
\label{sec:IESP}

This section summarizes the reduced-order electrochemical model used
in the SG-DKF. The model follows the SPMe-type framework
\citep{li2018parameter,li2019electrochemical} and is
implemented in discrete time with sampling period $\Delta t$.

Solid-phase diffusion in each electrode is approximated using the
parabolic method \citep{guo2025comparative,li2018parameter}. The surface stoichiometries are
\begin{align}
x_{s,p,\mathrm{surf}}(k)
  &= x_{s,p,\mathrm{avg}}(k) + \Delta x_{s,p}(k),\\
x_{s,n,\mathrm{surf}}(k)
  &= x_{s,n,\mathrm{avg}}(k) + \Delta x_{s,n}(k),
\end{align}
where $x_{s,p,\mathrm{avg}}(k)$ and $x_{s,n,\mathrm{avg}}(k)$ denote
the averaged stoichiometries and $\Delta x_{s,p}(k)$,
$\Delta x_{s,n}(k)$ are diffusion states.

Charge balance gives the averaged stoichiometry dynamics
\begin{align}
x_{s,p,\mathrm{avg}}(k+1)
  &= x_{s,p,\mathrm{avg}}(k)
     + \frac{D_p}{Q_{\mathrm{eff}}} I(k)\,\Delta t,
\label{eq:xsp_avg_disc}
\end{align}
\begin{align}
x_{s,n,\mathrm{avg}}(k+1)
  &= x_{s,n,\mathrm{avg}}(k) \notag\\
  &- \frac{D_n}{D_p}
       \bigl[x_{s,p,\mathrm{avg}}(k+1)-x_{s,p,\mathrm{avg}}(k)\bigr],
\label{eq:xsn_avg_disc}
\end{align}

where $D_p$ and $D_n$ are capacity-related coefficients. The parabolic
approximation yields first-order diffusion dynamics
\begin{align}
\Delta x_{s,p}(k+1) &= a_{p}\,\Delta x_{s,p}(k) + b_{p}\,I(k),\\
\Delta x_{s,n}(k+1) &= a_{n}\,\Delta x_{s,n}(k) + b_{n}\,I(k),
\end{align}
with $a_p,a_n,b_p,b_n$ functions of
$\tau_p^s,\tau_n^s,D_p,D_n,Q_{\mathrm{eff}}$
as detailed in \citep{li2018parameter}.

Rate-dependent usable capacity is described by Peukert’s law
\citep{yang2019comprehensive}:
\begin{equation}
Q_{\mathrm{eff}}
  = Q_{\mathrm{all}}
    \left(\frac{C_{\mathrm{ref}}}{C_{\mathrm{now}}}\right)^{n-1},
\end{equation}
where $Q_{\mathrm{all}}$ is the rated capacity,
$C_{\mathrm{ref}}$ the reference C-rate, $C_{\mathrm{now}}$ the
operating C-rate, and $n$ Peukert’s constant. The cell SOC keeps its
standard definition based on the rated capacity:
\begin{equation}
\mathrm{SOC}(k+1)
  = \mathrm{SOC}(k)
    - \frac{I(k)\,\Delta t}{Q_{\mathrm{all}}}.
\label{eq:soc_disc}
\end{equation}

Electrolyte diffusion is represented by two concentration-deviation
states at the positive and negative current collectors
\citep{li2018parameter}:
\begin{align}
\Delta c_1(k+1)
 &= \Delta c_1(k)
    + \frac{\Delta t}{\tau_e}
      \bigl[P_{\mathrm{con},a} I(k) - \Delta c_1(k)\bigr],\\
\Delta c_2(k+1)
 &= \Delta c_2(k)
    + \frac{\Delta t}{\tau_e}
      \bigl[P_{\mathrm{con},b} I(k) - \Delta c_2(k)\bigr],
\end{align}
where $\tau_e$ is the electrolyte diffusion time constant and
$P_{\mathrm{con},a}$, $P_{\mathrm{con},b}$ are cell-dependent
coefficients.

The corresponding concentration overpotential is
\begin{equation}
\eta_{\mathrm{con}}(k)
 = \frac{2RT}{F}(1-t_+)
   \ln\!\left(
   \frac{c_0 + \Delta c_1(k)}
        {c_0 - \Delta c_2(k)}
   \right),
\end{equation}
with $R$ the gas constant, $T$ the temperature, $F$ the Faraday
constant, $t_+$ the transference number, and $c_0$ the nominal
electrolyte concentration.

The reaction overpotential is obtained from Butler--Volmer kinetics:
\begin{align}
\eta_{\mathrm{rxn}}(k)
  &= \frac{2RT}{F}\Bigl[
       \ln\!\bigl(\sqrt{m_p^2(k)} + 1 + m_p(k)\bigr)
       \nonumber\\
  &\qquad\qquad\quad
       +\, \ln\!\bigl(\sqrt{m_n^2(k)} + 1 + m_n(k)\bigr)
     \Bigr].
\label{eq:eta_rxn}
\end{align}
where
\begin{align}
m_p(k) &=
 \frac{D_p P_{\mathrm{rxn},p}}
      {6 Q_{\mathrm{eff}} v_p^{1/2}
      \bigl(1-x_{s,p,\mathrm{surf}}(k)\bigr)^{1/2}}\, I(k),\\
m_n(k) &=
 \frac{D_n P_{\mathrm{rxn},n}}
      {6 Q_{\mathrm{eff}} v_n^{1/2}
      \bigl(1-x_{s,n,\mathrm{surf}}(k)\bigr)^{1/2}}\, I(k),
\end{align}
and $v_p,v_n,P_{\mathrm{rxn},p},P_{\mathrm{rxn},n}$ are
cell-specific constants.

The terminal voltage prediction is
\begin{align}
U(k)
  &= \bigl[\,U_p\!\bigl(x_{s,p,\mathrm{surf}}(k)\bigr)
           - U_n\!\bigl(x_{s,n,\mathrm{surf}}(k)\bigr)\bigr]
     \nonumber\\
  &\quad
     -\, \eta_{\mathrm{con}}(k)
     -\, \eta_{\mathrm{rxn}}(k)
     -\, R_{\mathrm{ohm}} I(k).
\label{eq:terminal_voltage}
\end{align}
The state EKF estimates the fast electrochemical dynamics using
\begin{equation}
x_k =
\begin{bmatrix}
\mathrm{SOC}(k),\;
\Delta x_{s,p}(k),\;
\Delta x_{s,n}(k),\;
\Delta c_1(k),\;
\Delta c_2(k)
\end{bmatrix}^{\!\top}.
\label{eq:state_vector}
\end{equation}

The parameter EKF tracks slowly varying electrochemical parameters:
\begin{equation}
\theta_k =
\begin{bmatrix}
D_p(k),\;
D_n(k),\;
Q_{\mathrm{all}}(k),\;
x_{s,p,0}(k),\;
x_{s,n,0}(k)
\end{bmatrix}^{\!\top}.
\label{eq:parameter_vector}
\end{equation}
where $x_{s,p,0}$ and $x_{s,n,0}$ are the initial stoichiometries of
the positive and negative electrodes. Collecting the above dynamics and
voltage expression yields the nonlinear model
$x_{k+1}=f(x_k,\theta_k,u_k)$, $y_k=h(x_k,\theta_k,u_k)$ used by the
SG-DKF in Section~\ref{sec:SGDKF}.

\section{Results and Discussion}

The estimation performance of the proposed SG-DKF method is evaluated under urban dynamometer driving schedule (UDDS) and 1C
constant-current discharge (C is a current rate, here 1C is 2.9A) at 25~\(^\circ\)C.  
Two scenarios are considered: (i) zero initial SOC error and (ii) 30\%
initial SOC error.  
The SG-DKF is compared with the conventional dual EKF. The experimental battery is a Panasonic NCR18650PF cell with a nominal capacity of 2.9~Ah. The positive electrode is composed of a Nickel–Manganese–Cobalt oxide (NMC), and the negative electrode uses graphite.

Fig.~\ref{fig:1} shows the SOC estimation results without initial SOC
error.  
Both algorithms accurately track the reference SOC under UDDS and 1C
conditions.  
The SG-DKF produces curves nearly identical to those of the dual EKF,
indicating that the dead-zone supervisor does not interfere with the
estimation when the initialization is reliable.  
Small RMSE reductions (Table~\ref{tab:1}) confirm that SG-DKF maintains the
nominal performance of the dual EKF while avoiding unnecessary parameter
corrections.

The benefits of the SG-DKF become more evident when a 30\% SOC initial error is
introduced, as shown in Fig.~\ref{fig:2}.  
The dual EKF exhibits long transient errors and fluctuations under both
profiles due to the strong coupling between state and parameter
corrections.  
In contrast, the SG-DKF converges significantly faster because the
dead-zone supervisor suppresses parameter updates during periods of high
innovation, allowing the state EKF to stabilize first.  
Once the innovation falls within the stability-guaranteed region, the
parameter EKF is reactivated and refines the model parameters without
destabilizing the filter.

Quantitatively, the SG-DKF reduces the RMSE from 2.56\% to 1.39\%
under UDDS and from 5.04\% to 2.73\% under 1C discharge, corresponding
to improvements of more than 45\%.  
These results demonstrate that the SG-DKF preserves accuracy in nominal
conditions while substantially enhancing robustness against poor SOC
initialization.

\begin{table}[H]
\centering
\caption{SOC estimation errors}
\label{tab:1}
\begin{tabular}{l c l c}
\toprule
\makecell{\textbf{Working}\\\textbf{condition}} 
& \makecell{\textbf{Initial}\\\textbf{SOC error / \%}}
& \textbf{Algorithm} 
& \textbf{RMSE / \%} \\
\midrule
UDDS & 0   & Dual EKF  & 0.19 \\
UDDS & 0   & DG-DKF    & 0.18 \\
\midrule
UDDS & 30  & Dual EKF  & 2.56 \\
UDDS & 30  & DG-DKF    & 1.39 \\
\midrule
1C   & 0   & Dual EKF  & 0.31 \\
1C   & 0   & DG-DKF    & 0.30 \\
\midrule
1C   & 30  & Dual EKF  & 5.04  \\
1C   & 30  & DG-DKF    & 2.73 \\
\bottomrule
\end{tabular}
\end{table}

\section{Conclusion}
This paper addressed the stability problem of dual Kalman filtering
and proposed a SG-DKF framework. The starting point was a general
dual EKF structure, in which one filter estimates the system states and
the other filter estimates slowly varying parameters. A Lyapunov
function was constructed for the state estimation error dynamics, and a
sufficient condition was derived that relates the innovation, the
Kalman gain and the state error to a stability bound. Based on this
analysis, an adaptive dead zone criterion was obtained that determines
whether the parameter filter should be updated or temporarily
suspended. The SG-DKF framework was then combined with an electrochemical model of a Panasonic NCR18650PF cell. The experimental results under UDDS and 1C conditions at 25~\(^\circ\)C showed that SG-DKF preserves the estimation accuracy of a conventional
dual EKF when the initial SOC is well specified. At the same time, it
significantly improves robustness against large initial SOC errors,
reducing the SOC RMSE by more than 45\% in the tested scenarios. These
results confirm that the dead zone supervisor effectively mitigates the
adverse effects of state parameter coupling and prevents divergence of
the dual estimator. Future work will extend the proposed framework to incorporate temperature dependent electrochemical models and ageing related parameter evolution in order to handle wider operating windows and long term degradation. 
\begin{ack}
This work was supported by the Research Foundation - Flanders (FWO) (grant numbers 1252326N).
\end{ack}

\bibliography{ifacconf}             
                                                   







\end{document}